\documentclass[showpacs,preprintnumbers,amsmath,amssymb,aps,prl]{revtex4}
\usepackage{graphicx}
\usepackage{mathptmx}      
\begin{document}

\title{On the Gaussian approximation for master equations
}


\author{Luis F. Lafuerza, Raul Toral 
}


\affiliation{IFSIC (Instituto de F{\'\i}sica Interdisciplinar y Sistemas Complejos)\\ CSIC-University of the Balearic Islands,
	Campus UIB, 07122-Palma de Mallorca, Spain}

\begin{abstract}
We analyze the Gaussian approximation as a method to obtain the first and second moments of a stochastic process described by a master equation. We justify the use of this approximation with ideas coming from van Kampen's expansion approach (the fact that the probability distribution is Gaussian at first order). We analyze the scaling of the error with a large parameter of the system and compare it with van Kampen's method. Our theoretical analysis and the study of several examples shows that the Gaussian approximation turns out to be more accurate. This could be specially important for problems involving stochastic processes in systems with a small number of particles.

\end{abstract}
\date{\today}

\maketitle

\section{Introduction}
\label{intro}

Master equations are a convenient tool to treat stochastic Markov processes\cite{VK,Gardiner}. In some cases, they offer an alternative approach to the Chapman-Kolmogorov equation and have been used extensively in discrete-jumps, or birth-death, processes, such as chemical reactions (including those happening inside a cell), population dynamics or other ecology problems\cite{Gillespie}, opinion formation and cultural transmission in the field of sociophysics\cite{castellano}, etc. In all these cases, it is important to consider that the population number (whether molecules, individuals, agents, etc.) might not be very large (maybe ranging in the tens or hundreds) and the fluctuations, that typically scale as the square root of the inverse of this number, can not be considered as negligible. It is therefore, of the greatest importance to derive evolution equations for the average behavior and the fluctuations. The important work by van Kampen\cite{VK} offers a systematic way of deriving these equations from an expansion of the master equation in a parameter $\Omega$, typically the system volume. The $\Omega$-expansion is mostly used in its lowest order form, in which one can prove that the error in the average value, the second moment and the fluctuations (the variance), scale at most as $\Omega^{0}$, $\Omega^{1}$ and $\Omega^{1/2}$, respectively. The van Kampen $\Omega$-expansion, furthermore, shows that, at this lowest order, the fluctuations follow a Gaussian distribution. In this paper, we take this result of van Kampen's theory and, considering from the very beginning that fluctuations are Gaussian, we derive a closed system of equations for the average value and the second moment.  This Gaussian closure of the hierarchy of moments turns out to be more accurate that the $\Omega$-expansion as the above-mentioned errors scale at most as $\Omega^{-1/2}$, $\Omega^{1/2}$ and $\Omega^{1/2}$, respectively. Furthermore, the Gaussian closure scheme is very simple to carry on in practice and can be easily generalized to systems described by more than one variable.

The paper is organized as follows:  In the following section, we will briefly review the $\Omega$-expansion and derive the main equations for the Gaussian closure approximation. The errors of both methods are discussed in section \ref{sec:error}. In sections \ref{sec:annihilation} and \ref{sec:autocatalytic}, we will give examples of the application of the method in the cases of a binary chemical reaction and an autocatalytic reaction. The results of these two examples confirm the error-analysis performed before. For both processes we compare with the results coming from the exact solution of the master equation in the stationary regime (derived in the appendix for the binary chemical reaction), and the results of numerical simulations using the Gillespie algorithm in the time-dependent evolution. In section \ref{sec:opinion} we present an application to a recently introduced model for opinion formation which requires two variables for its full description. Finally, in section \ref{sec:conclusions} we end with a brief summary of the work.

\section[Formulation]{Formulation}

Let $P(n,t)$ be the probability that at time $t$ the population number takes the value $n$. We consider that it evolves according to a general master equation of the form:
\begin{equation}
\frac{\partial P(n,t)}{\partial t}=\sum_{k}(E^{k}-1)\left[C_{k}(n;\Omega)P(n,t)\right],
 \label{ME}
\end{equation} 
where $E$ is the linear step operator such that $E^k[f(n)]\equiv f(n+k)$ and $k$ runs over the integer numbers. Besides  $n$, the coefficients $C_k(n;\Omega)$ depend on $\Omega$, which is a large parameter of the system (typically the system volume). We consider that these functions are polynomials or can be expanded in power series of $n$ as $C_k(n;\Omega)=\sum_a C_k^a(\Omega)n^a$ where  the coefficients $C_k^a(\Omega)$ scale as \\$C_k^a(\Omega)=\Omega^{1-a}\left(c^a_{k,0}+c^{a}_{k,1}\Omega^{-1}+c^{a}_{k,2}\Omega^{-2}+\dots\right)$. Master equations of this form appear in the description of chemical reactions \cite{Gillespie}, ecological systems \cite{Pigolotti}  and opinion dynamics \cite{Wio}, among many other cases. More specific examples will be considered in the next sections.

In a seminal work, van Kampen\cite{VK} has given a way of finding an approximate solution of Eq. (\ref{ME}). The approximation is based upon the splitting of the variable $n$ using the ansatz $n=\Omega\phi(t)+\Omega^{\frac{1}{2}}\xi$, where $\phi(t)\sim O(\Omega^{0})$ is a function of time accounting for the deterministic part of $n$ and $\xi\sim O(\Omega^{0})$ corresponds to the fluctuations. Changing variables from $n$ to $\xi$ in Eq. (\ref{ME}), and expanding in powers of $\Omega$ one obtains a Fokker-Planck equation for the probability distribution $\Pi(\xi,t)$ of the new variable $\xi$:
\begin{equation}
\label{fpe}
 \frac{\partial\Pi(\xi,t)}{\partial t}=\left[\sum_{a,k}c_{k,0}^{a}ka\phi^{a-1}\right]\frac{\partial(\xi\Pi)}{\partial\xi}+\left[\sum_{a,k}c_{k,0}^{a}\frac{k^{2}}{2}\phi^{a}\right]\frac{\partial^{2}\Pi}{\partial\xi^{2}}+O(\Omega^{-\frac{1}{2}}),
\end{equation}
where the macroscopic variable $\phi$ satisfies
\begin{equation}
\label{macro}
 \frac{d\phi(t)}{dt}=\sum_{a,k}kc_{k,0}^{a}\phi^{a}.
\end{equation}
From Eq.(\ref{fpe}) we obtain the first and second moments of the fluctuations:
\begin{eqnarray}
\frac{\partial\langle\xi\rangle}{\partial t}&=&-\left[\sum_{a,k}c_{k,0}^{a}ka\phi^{a-1}\right]\langle \xi\rangle, \label{vk2} \\
\frac{\partial\langle\xi^{2}\rangle}{\partial t}&=&-2\left[\sum_{a,k}c_{k,0}^{a}ka\phi^{a-1}\right]\langle \xi^{2}\rangle+2\left[\sum_{a,k}c_{k,0}^{a}\frac{k^{2}}{2}\phi^{a}\right].\label{vk3}
\end{eqnarray}

As proven by van Kampen, the solution of the Fokker-Planck equation (\ref{fpe}) is a Gaussian distribution. Therefore, the $\Omega$-expansion method tells us that, up to corrections of order $\Omega^{-\frac{1}{2}}$, the fluctuations of the variable $n$ follow a Gaussian distribution. It suffices, then, to know the first and second moments of this distribution. Our intention is to use from the very beginning the Gaussian property in order to obtain a closed system of equations for the first two moments $\langle n\rangle$ and $\langle n^2\rangle$.

From (\ref{ME}) we get the (exact) equations for these first two moments, as:
\begin{equation}
\frac{d\langle n\rangle}{dt}=-\sum_{k}\left\langle kC_k(n;\Omega)\right\rangle,\label{dndt}\hspace{1.0cm}
\frac{d\langle n^{2}\rangle}{dt}=\sum_{k}\left\langle k(k-n) C_k(n;\Omega)\right\rangle.
\end{equation} 
After substitution of the series expansion $C_k(n;\Omega)=\sum_a C_k^a(\Omega)n^a$ in the right hand side of these equations,  one obtains higher order moments $\langle n^m\rangle$ for $m\ge 3$. The Gaussian closure replaces these higher order moments with the expressions $\langle n^m\rangle_G$ that hold in the case of a Gaussian distribution, i.e. $\langle n\rangle_G=\langle n\rangle$, $\langle n^2\rangle_G=\langle n^2\rangle$ and 
\begin{equation}
\langle n^m\rangle_G=\langle n\rangle^m+\sum_{k=1}^{\left[\frac{m}{2}\right]}{m \choose 2k} (2k-1)!!\langle n\rangle^{m-2k}\left[\langle n^2\rangle -\langle n\rangle^2\right]^k
\end{equation}
for $m\ge 3$. The first moments are explicitly shown in table \ref{gaussmoments}.

\begin{table}[h c]
\begin{tabular}{| c | l |}
\hline
 Moment & Gaussian approximation\\
\hline
$\langle n^{3}\rangle$& $3\langle n^{2}\rangle\langle n\rangle-2\langle n\rangle^{3}$ \\
\hline
$\langle n^{4}\rangle$ & $3\langle n^{2}\rangle^{2}-2\langle n\rangle^{4}$\\ 
\hline
$\langle n^{5}\rangle$ & $15\langle n^{2}\rangle^{2}\langle n\rangle-20\langle n^{2}\rangle\langle n\rangle^{3}+6\langle n\rangle^{5}$ \\
\hline
$\langle n^{6}\rangle$ & $15\langle n^{2}\rangle^{3}-30\langle n^{2}\rangle\langle n\rangle^{4}+45\langle n\rangle^{6}$ \\
\hline
$\langle n_{1}^{2}n_{2}\rangle$ & $\langle n_{1}^{2}\rangle\langle n_{2}\rangle+2\langle n_{1}\rangle\langle n_{1}n_{2}\rangle-2\langle n_{1}\rangle^{2}\langle n_{2}\rangle$\\
\hline
$\langle n_{1}^{2}n_{2}^{2}\rangle$ & $\langle n_{1}^{2}\rangle\langle n_{2}^{2}\rangle+2\langle n_{1}n_{2}\rangle^{2}-2\langle n{1}\rangle^{2}\langle n{2}\rangle^{2}$\\
\hline
$\langle n_{1}^{3}n_{2}\rangle$ & $3\langle n_{1}^{2}\rangle\langle n_{1}n_{2}\rangle-2\langle n{1}\rangle^{3}\langle n{2}\rangle$\\
\hline
$\langle n_{1}^{3}n_{2}^{2}\rangle$ & $ \begin{array}{l}6\langle n_{1}n_{2}\rangle^{2}\langle n_{1}\rangle+6\langle n{1}\rangle^{3}\langle n{2}\rangle^{2}+6\langle n_{1}n{2}\rangle\langle n_{2}\rangle(\langle n_{1}\rangle^{2}-2\langle n_{1}\rangle^{2}\cr
-6\langle n_{1}^{2}\rangle\langle n_{2}\rangle^{2}\langle n_{1}\rangle+3\langle n_{1}^{2}\rangle\langle n_{2}^{2}\rangle\langle n_{1}\rangle-2\langle n_{1}\rangle^{3}\langle n_{2}^{2}\rangle\end{array}$\\
\hline
\end{tabular}
\caption{Gaussian moments}
\label{gaussmoments}
\end{table}

The van Kampen ansatz $n=\Omega\phi(t)+\Omega^{\frac{1}{2}}\xi$ allows us to find the error of this approximation. It follows that:
\begin{equation}
\frac{\langle n^m\rangle}{\Omega^{m-1}}=\sum_{l=0}^m{m \choose l}\Omega^{1-l/2}\phi^{m-l}\langle \xi^l\rangle.
\end{equation}
In the Gaussian approximation, the first three terms of the sum, $l=0,1,2$ are exact and the term $l=3$ scales as $\Omega^{-1/2}$, or:
\begin{equation}
\label{errg}
\frac{\langle n^m\rangle}{\Omega^{m-1}}=\frac{\langle n^m\rangle_G}{\Omega^{m-1}}+O(\Omega^{-1/2}).
\end{equation}
If we use this result in each of the terms of Eq.(\ref{dndt}) and $C_k(n;\Omega)=\Omega^{1-a}(c_{k,0}^a+O(\Omega^{-1}))$ we obtain
\begin{equation}
\label{dndtf}
\frac{d\langle n\rangle}{dt}=g_1(\langle n \rangle,\langle n^2 \rangle)+O(\Omega^{-1/2}),
\end{equation}
with $g_1\equiv \displaystyle-\sum_{k}\left\langle kC_k(n;\Omega)\right\rangle_G$.
Similarly, one finds
\begin{equation}
\label{dn2dtf}
\frac{d\langle n^2\rangle}{dt}=g_2(\langle n \rangle,\langle n^2 \rangle)+O(\Omega^{1/2}),
\end{equation}
with $g_2\equiv\displaystyle\sum_{k}\left\langle k(k-n) C_k(n;\Omega)\right\rangle_G$.

This Gaussian approximation scheme (or equivalently, finding a hierarchy of equations for the cumulants and neglecting those of order greater than two) has been used many times in the literature in different contexts \cite{Zwanzig,Cubero}. We will show in the next section that the direct use of Eqs. (\ref{dndtf},\ref{dn2dtf}) has a smaller error that the use of Eqs. (\ref{macro}-\ref{vk3}). Before showing this, we will generalize this procedure for the case of two-variable problems. Let us consider a master equation of the following form:
\begin{equation}
\frac{\partial P(n_{1},n_{2},t)}{\partial t}= \sum_{k_{1},k_{2}}(E^{k_{1}}_{1}E^{k_{2}}_{2}-1)\left[C_{k_1,k_2}(n_1,n_2;\Omega)P(n_{1},n_{2},t)\right].
 \label{ME2v}
\end{equation} 

The evolution equations for the first, second order moments and the correlations are:
\begin{eqnarray}
\frac{d\langle n_{i}\rangle}{dt}&=-&\sum_{k_1,k_2}\left\langle k_iC_{k_1,k_2}(n_1,n_2;\Omega)\right\rangle,\label{dndt21}\\
\frac{d\langle n_{i}^{2}\rangle}{dt}&=&\sum_{k_1,k_2}\left\langle k_i(k_i-n_i) C_{k_1,k_2}(n_1,n_2;\Omega)\right\rangle,\label{dndt22}\\
\frac{d\langle n_{1}n_{2}\rangle}{dt}&=&\sum_{k_1,k_2}\left\langle (k_{1}k_{2}-k_{2}n_{1}-k_{1}n_{2}) C_{k_1,k_2}(n_1,n_2;\Omega)\right\rangle,\label{dndt23}
\end{eqnarray}
($i=1,2$). Again, the Gaussian closure consists in replacing $\langle n_1^{m_1}n_2^{m_2}\rangle$ by the expression $\langle n_1^{m_1}n_2^{m_2}\rangle_G$ that holds assuming that the joint distribution $P(n_{1},n_{2},t)$ is Gaussian. This can be computed using Wick's theorem\cite{wick}. In table (\ref{gaussmoments}) we write the expression of some of the terms.

\section[Error of the method]{Error of the method}
\label{sec:error}
We now calculate the error of the Gaussian approximation and compare it with the one of the $\Omega$-expansion. In Eqs. (\ref{dndtf}-\ref{dn2dtf}) we have shown that the errors we introduce in the equations for the moments when performing the Gaussian approximation are of order $O(\Omega^{-1/2})$ for $\langle n\rangle$ and $O(\Omega^{1/2})$ for $\langle n^{2}\rangle$. The Gaussian approximation scheme proceeds by considering approximations $\mu_1(t)$, $\mu_2(t)$ to the true moments $\langle n(t)\rangle$, $\langle n^2(t) \rangle$. These approximations are defined as the solution of the evolution equations (\ref{dndtf},\ref{dn2dtf}):
\begin{equation}
\frac{d\mu_1}{dt}=g_{1}(\mu_1,\mu_2), \label{f1g}\hspace{1.0cm}
\frac{d\mu_2}{dt}=g_{2}(\mu_1,\mu_2).
\end{equation}
Defining the errors $\epsilon_1,\epsilon_2$ as: $\langle n\rangle=\mu_1+\epsilon_1$, $\langle n^{2}\rangle=\mu_2+\epsilon_2$; expanding in first order in $\epsilon_1$ and $\epsilon_2$, and using equations (\ref{dndtf}-\ref{dn2dtf}) and (\ref{f1g}) we get:
\begin{eqnarray}
\frac{d\epsilon_1}{dt}&=&\frac{\partial g_{1}(\mu_1,\mu_2)}{\partial \mu_1}\epsilon_1+\frac{\partial g_{1}(\mu_1,\mu_2)}{\partial\mu_2}\epsilon_2+O(\Omega^{-1/2}),\\
\frac{d\epsilon_2}{dt}&=&\frac{\partial g_{2}(\mu_1,\mu_2)}{\partial \mu_1}\epsilon_1+\frac{\partial g_{2}(\mu_1,\mu_2)}{\partial\mu_2}\epsilon_2+O(\Omega^{1/2}).
\end{eqnarray}
Taking into account that $\mu_1,g_{1}\sim O(\Omega)$, $\mu_2,g_{2}\sim O(\Omega^{2})$, we have:
\begin{eqnarray}
\frac{d\epsilon_1}{dt}&=&O(\Omega^{0})\epsilon_1+O(\Omega^{-1})\epsilon_2+O(\Omega^{-1/2}), \label{depsilondt} \\
\frac{d\epsilon_2}{dt}&=&O(\Omega)\epsilon_1+O(\Omega^{0})\epsilon_2+O(\Omega^{1/2}).\label{depsilon2dt}
\end{eqnarray}
If we set $\epsilon_1\sim O(\Omega^{a})$, $\epsilon_2\sim O(\Omega^{b})$, and the initial conditions are known, so that initially $\epsilon_1=\epsilon_2=0$, equations (\ref{depsilondt}), (\ref{depsilon2dt}) imply that $a\le -1/2$ and $b\le 1/2$, a scaling respected during the time evolution.

In conclusion, solving equations (\ref{dndtf}-\ref{dn2dtf}), we get $\langle n\rangle$ and $\langle n^{2}\rangle$ with errors of order $\epsilon_1=O(\Omega^{-1/2})$ and  $\epsilon_2=O(\Omega^{1/2})$, or smaller. Using the equations (\ref{macro}-\ref{vk3}) of first order van Kampen's expansion the error is of higher order in both cases: $O(\Omega^{0})$ for $\langle n\rangle$ and $O(\Omega^{1})$ for $\langle n^{2}\rangle$. However, for the variance, $\sigma^{2}\equiv\langle n^{2}\rangle-\langle n\rangle^{2}$, both approximations have an error of order $O(\Omega^{1/2})$. We will show in the next sections that the Gaussian approximation has the extra advantage that it is easier to derive for many problems of practical interest. 

One might be tempted to go to higher order schemes, where one neglects all the cumulants of order greater than $m$ with $m>2$, and in this way obtain a closed set of equations for the first $m$ moments. For example, if we neglect all the cumulants of order greater than 3, applying the same analysis as before, it is possible to derive that the errors in the first, second and third moments are of order $O(\Omega^{-1},\Omega^0,\Omega^{1})$, respectively. 

A word of caution is needed here. When truncating beyond the second cumulant, it is not ensured that the resulting probability distribution is positive definite \cite{Hanggi}. This means that one could get from such an scheme inconsistent results, e.g. a negative variance. Nevertheless, according to our analysis, the importance of these spurious results would decrease with $\Omega$ as indicated, so one can still get useful results from higher order schemes. 

In the following sections we will compare the Gaussian approximation presented here with the first order $\Omega$-expansion in some specific examples.

\section{Binary reaction $A +B {{\kappa \atop \longrightarrow}\atop{\longleftarrow \atop \omega}} C$}
\label{sec:annihilation} 

Chemical reactions are suitable processes for a stochastic description. The stochastic approach is specially necessary when the number of molecules considered is small, as it is the frequently addressed case of chemical reactions inside a cell, because in this situation fluctuations can be very important.

We consider the general process $A +B {{\kappa \atop \longrightarrow}\atop{\longleftarrow \atop \omega}} C$, limited by reaction. This means that any two particles $A$ and $B$ have the same probability of reaction. Denoting by $A(t)$ and $B(t)$, respectively, the number of molecules of the $A$ and $B$ substances, the rate for the $A+B\longrightarrow C$ reaction is $\frac{\kappa}{\Omega}A(t)B(t)$. For the reverse reaction, it is assumed that $C$ has a constant concentration, and hence the rate is $\omega  \Omega$. In these expressions $\Omega$ is proportional to the total volume accessible. Since $B(t)-A(t)\equiv \Delta$ is a constant, one only needs to consider one variable, for example, the number of $A$ molecules at time $t$. Let us denote by $P(n,t)$ the probability that there are $n$ $A$-molecules at time $t$. The master equation describing the process is:
\begin{equation}
\label{eq:master}
\frac{dP(n,t)}{dt}= \frac{\kappa}{\Omega}\left[(n+1)(\Delta +n+1)P(n+1,t)-n(n+\Delta)P(n,t)\right]
+\omega\Omega[P(n-1,t)-P(n,t)],
\end{equation}
which is the basis of the subsequent analysis. Note that this equation can be written in the form (\ref{ME}) setting $C_{1}(n;\Omega)=\frac{\kappa}{\Omega}n(n+\Delta),C_{-1}(n;\Omega)=\omega\Omega$.

In the irreversible case, $\omega=0$, this master equation can be solved exactly using the generating function technique. In the general case, $\omega\neq0$, an exact solution can also be found for the stationary state $\frac{\partial P(n,t)}{\partial t}=0$. Details of the calculation are given in the appendix. We will compare the results obtained from the Gaussian approximation and the first order $\Omega$-expansion with the exact results, when available. 

The equations for the first two moments, using (\ref{dndt}), are:
\begin{eqnarray}
\frac{d\langle n\rangle}{dt}&=&-\frac{\kappa}{\Omega}\left(\langle n^2\rangle +\Delta\langle n\rangle\right)+\Omega\omega,\label{dndtd}\\
\frac{d\langle n^{2}\rangle}{dt}&=&\frac{\kappa}{\Omega}(-2\langle n^3\rangle +(1-2\Delta)\langle n^2\rangle+\Delta\langle n\rangle)-2\Omega\omega\langle n\rangle+\Omega\omega.
 \label{dn2dtd}
\end{eqnarray}

Using the Gaussian approximation, the evolution equations for the moments are:
\begin{eqnarray}
\frac{d\mu_1}{dt}&=&-\frac{\kappa}{\Omega}(\mu_2+\Delta\mu_1)+\Omega\omega,\label{ngreac}\\
\frac{\mu_2}{dt}&=&\frac{\kappa}{\Omega}(4\mu_1^{3}-6\mu_2\mu_1+(1-2\Delta)\mu_2+\Delta\mu_1)+2\Omega\omega\mu_{1}+\Omega\omega.\label{n2greac}
\end{eqnarray}

And the first order $\Omega$-expansion gives:
\begin{eqnarray}
\frac{d \phi}{dt}&=&-\kappa\phi(\phi+\delta)+\omega,\label{n1vkreac}\\
\frac{d\langle n\rangle}{dt}&=&\kappa\Omega\phi^{2}-\kappa(\delta+2\phi)\langle n\rangle+\Omega\omega,\label{n2vkreac}\\
\frac{d\langle n^{2}\rangle}{dt}&=&-2\kappa(2\phi+\delta)\langle n^{2}\rangle+\Omega\left[\kappa\phi(\phi+\delta)(1-2\langle n\rangle)\right.+\cr
&& \left.2(\kappa\Omega\phi^{2}(2\phi+\delta)+\omega(\langle n\rangle+1+\omega\phi))\right],\label{n3vkreac}
\end{eqnarray}
where $\delta=\frac{\Delta}{\Omega}$.

We compare the two approximations in the time-dependent case with results obtained by averaging over single realizations of the process, obtained numerically using the Gillespie algorithm\cite{Gillespie}. In the next figures we compare the exact results with those obtained from the Gaussian approximation (computed by numerical integration of equations \ref{ngreac}, \ref{n2greac}) and  $\Omega$-expansion (equations \ref{n1vkreac}-\ref{n3vkreac}).

\begin{figure}[h]
\centering
\includegraphics[scale=0.5,angle=0,clip]{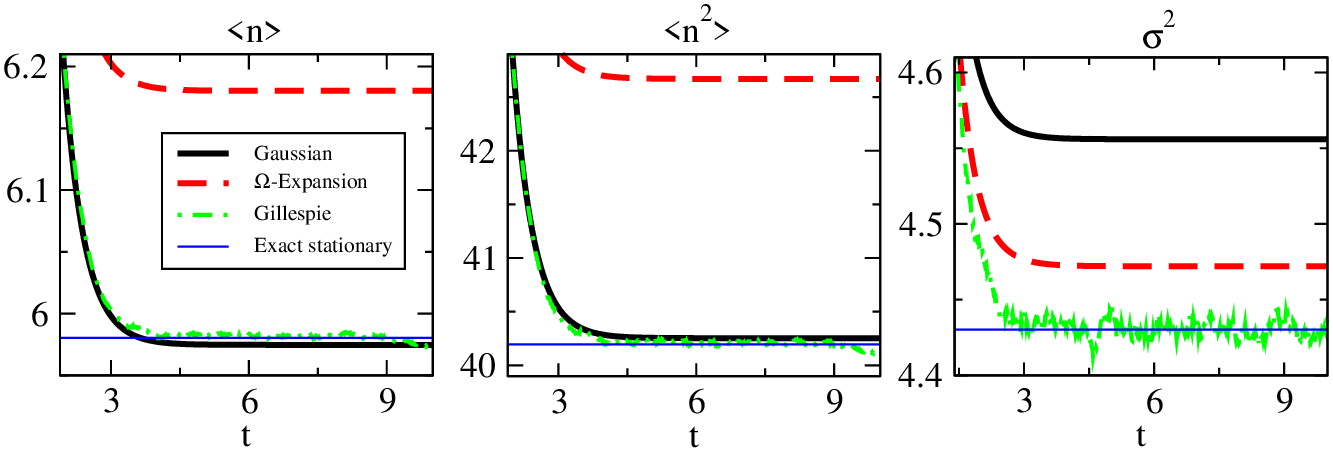}
\caption{$\langle n(t)\rangle$, $\langle n^{2}(t)\rangle$ and $\sigma^{2}(t)$ for the binary reaction $A +B {{\kappa \atop \longrightarrow}\atop{\longleftarrow \atop \omega}} C$ with parameters $\kappa=1$, $\omega=1$, $\Omega=10$ and initial conditions  $n(0)=100$, $\delta=1$. For the first two moments the Gaussian approximation (solid) is very close to the results obtained with the Gillespie algorithm (dot-dashed, obtained averaging over one million realizations) and the exact stationary value (thin line), while $1^{st}$ order $\Omega$-expansion (dashed) gives clearly different values. For $\sigma^{2}$, the $\Omega$-expansion gives more accurate results but both approximations differ from the exact values.} \label{anieq}
\end{figure}

\begin{figure}[h]
\centering
\includegraphics[scale=0.5,angle=0,clip]{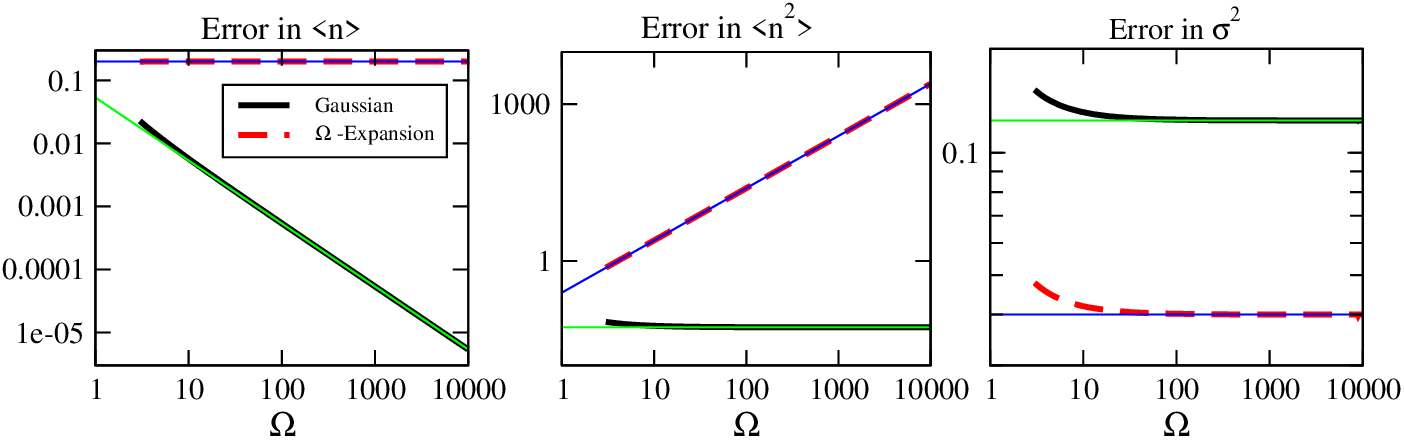}
\caption{Error in $\langle n\rangle$, $\langle n^{2}\rangle$ and $\sigma^{2}$ in the stationary state in the same case than in Fig.\ref{anieq}. The straight thin lines are fits to the data and have slope $-1$, $0$ or $1$. For the Gaussian approximation (solid), the errors in $(\langle n\rangle,\,\langle n^{2}\rangle,\, \sigma^{2})$ scale as $(\Omega^{-1},\,\Omega^{0},\,\Omega^{0})$. For the $\Omega$-expansion (dashed), the errors scale as $(\Omega^{0},\,\Omega^{1},\,\Omega^{0})$.} \label{aniesc}
\end{figure}

Figure (\ref{anieq}) shows that the Gaussian approximation reproduces better the exact results for the first two moments; for the variance, the $\Omega$-expansion gives more accurate results but both approximations differ from the exact values. Figure (\ref{aniesc}) shows that the errors in the stationary state, coming from the Gaussian approximation for the mean value, the second moment and the variance scale as $(\Omega^{-1},\,\Omega^{0},\,\Omega^0)$, respectively, while the errors of the $\Omega$-expansion at  first order scale as $(\Omega^{0},\,\Omega^{1},\,\Omega^{0})$. This scaling is consistent with the previous analysis, as the exponents of the errors are smaller than the obtained bounds.

\section{Autocatalytic reaction $A {{k \atop \longrightarrow}\atop{}} X$, $2X {{k' \atop \longrightarrow}\atop{}} B$}
\label{sec:autocatalytic}
The master equation describing this process is\cite{VK}:
\begin{equation}\frac{\partial P(n,t)}{\partial t}=\Omega\phi_{A}k[P(n-1,t)-P(n,t)]+\frac{k'}{\Omega}[(n+2)(n+1)P(n+2,t)-n(n-1)P(n,t)], \label{ME2}
\end{equation} 
where the concentration of $A$ particles is consider to be constant with a value $\phi_{A}$. This equation if of the form (\ref{ME}) with $C_{-1}(n;\Omega)=\Omega k\phi_{A}, C_{2}(n;\Omega)+\frac{k'}{\Omega}n(n-1)$. The general solution for this equation is not known, but the stationary solution $P^{st}(n)$ can be obtained using the generating function technique\cite{VK}. The exact equations for the first moments are:
\begin{eqnarray}
 \frac{d\langle n\rangle}{dt}&=&\Omega k\phi_{A}+2k'\frac{\langle n\rangle}{\Omega}-2k'\frac{\langle n^{2}\rangle}{\Omega},\\
\frac{d\langle n^{2}\rangle}{dt}&=&\Omega k\phi_{A}(2\langle n\rangle+1)-\frac{k'}{\Omega}(4\langle n^{3}\rangle-8\langle n^{2}\rangle+4\langle n\rangle).
\end{eqnarray}

Performing the Gaussian approximation, we get:
\begin{eqnarray}
 \frac{d\mu_1}{dt}&=&\Omega k\phi_{A}+2k'\frac{\mu_1}{\Omega}-2k'\frac{\mu_2}{\Omega},\label{ngauto}\\
 \frac{d\mu_2}{dt}&=&\Omega k\phi_{A}(2\mu_1+1)-\frac{k'}{\Omega}(12\mu_2\mu_1-8\mu_1^{3}-8\mu_2+4\mu_1).\label{n2gauto}
\end{eqnarray}

While first order $\Omega$-expansion approach leads to:
\begin{eqnarray}
\frac{d\phi}{dt}&=&k\phi{A}-2k'\phi^{2},\label{n1vkauto}\\
\frac{d\langle n\rangle}{dt}&=&\Omega(k\phi_{A}+2k'\phi^{2})-4k'\phi\langle n\rangle,\label{n2vkauto}\\
\frac{d\langle n^{2}\rangle}{dt}&=&-8k'\phi\langle n^{2}\rangle+\Omega(2k\phi_{A}+4k'\phi^{2})\langle n\rangle+\Omega(k\phi_{A}+4k'\phi^{2}). \label{n3vkauto}
\end{eqnarray}

In the next figures we show the results obtained with the Gaussian approximation (computed by numerical integration of equations \ref{ngauto}-\ref{n2gauto}), $\Omega$-expansion (equations \ref{n1vkauto}-\ref{n3vkauto}), the Gillespie algorithm, and the exact stationary solution.
\begin{figure}[h]
\centering
\includegraphics[scale=0.5,angle=0,clip]{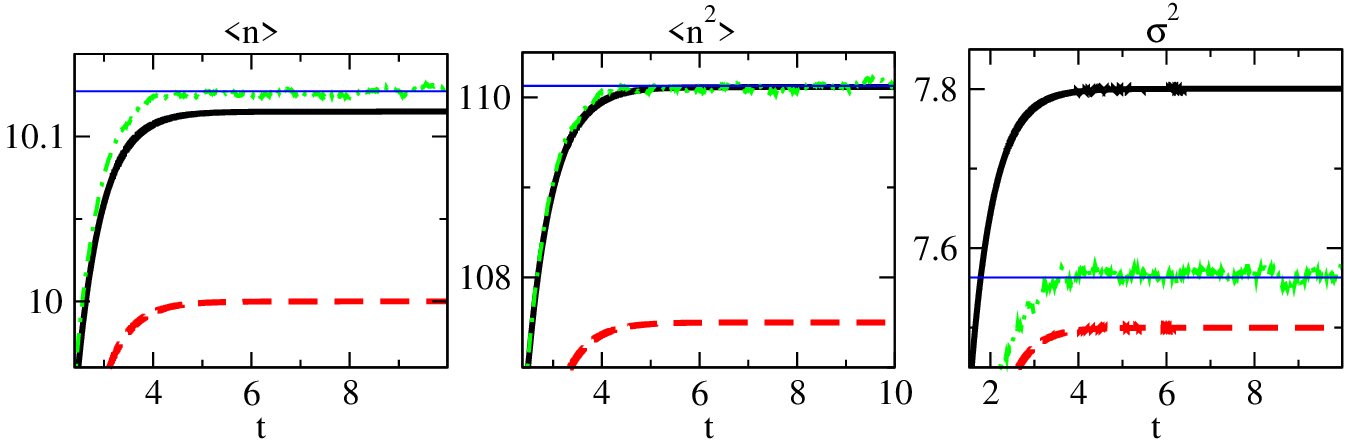}
\caption{$\langle n(t)\rangle$, $\langle n(t)^{2}\rangle$ and $\sigma^{2}(t)$ for the autocatalytic reaction $A {{k \atop \longrightarrow}\atop{}} X$, $2X {{k' \atop \longrightarrow}\atop{}} B$ with $k\phi_{A}=1$, $k'=1/2$, $\Omega=10$ and initial condition $n(0)=0$. For the first two moments the Gaussian approximation (solid) is very close to the results coming from the Gillespie algorithm (dot-dashed) and the exact value in the stationary case (thin line) whereas the $\Omega$-expansion result (dashed) is clearly different, although for $\sigma^{2}$ the $\Omega$-expansion provides more accurate results.\label{autocat}} 
\end{figure}

\begin{figure}[h]
\centering
\includegraphics[scale=0.5,angle=0,clip]{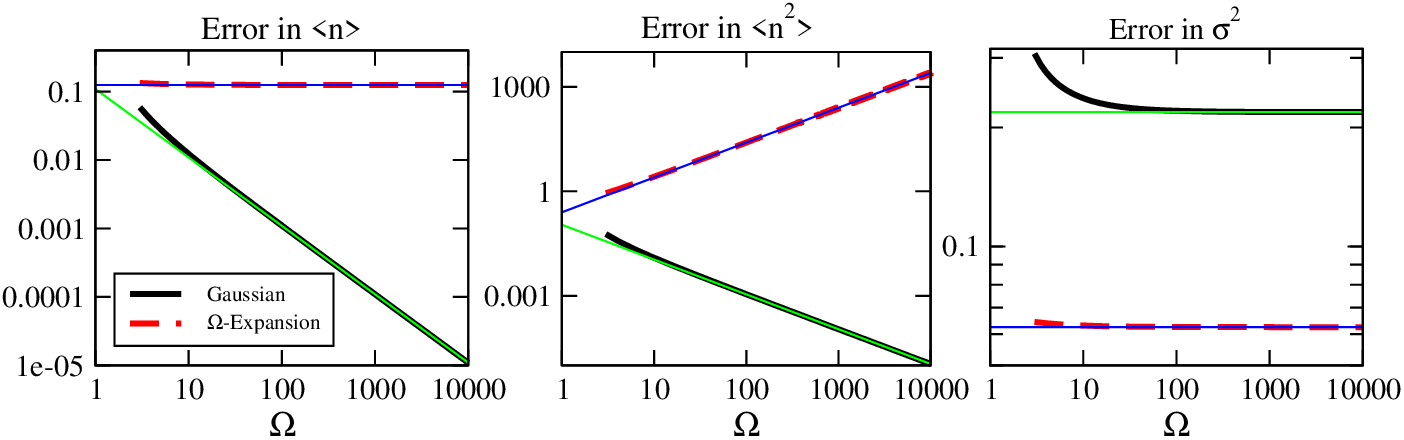}
\caption{Error in $\langle n\rangle,\langle n^{2}\rangle$ and $\sigma^{2}$ in the stationary state as a function of $\Omega$ in the same case than in Fig.\ref{autocat}. The thin lines have slope $-1$, $0$ or $1$. For the Gaussian approximation (solid), the errors in $(\langle n\rangle, \langle n^{2}\rangle,\sigma^{2})$ scale (asymptotically) as $(\Omega^{-1},\,\Omega^{-1},\,\Omega^{0})$. For the $\Omega$-expansion, the errors scale as $(\Omega^{0},\,\Omega^{1},\,\Omega^{0})$.}
\label{scalingerr} 
\end{figure}

As in the previous example, we see that the Gaussian approximation fits better the evolution of the moments, but the variance is somehow better approximated by the first order $\Omega$-expansion. In figure (\ref{scalingerr}) we show the errors in the stationary state for the two approximations as a function of  $\Omega$. We see that the errors in $(\langle n\rangle,\,\langle n^{2}\rangle,\,\sigma^2)$ decay as $(\Omega^{-1},\,\Omega^{-1},\Omega^0)$ for the Gaussian approximation, while the first-order $\Omega$-expansion leads to errors that scale as $(\Omega^0,\,\Omega^1,\Omega^0)$. Again, this scaling is consistent with the analysis of the approximations performed.

\section{Opinion formation}
\label{sec:opinion}
In the last few years there has been a growing interest in the application of methods and techniques coming from statistical physics to the study of complex phenomena in fields traditionally far from physics research, particularly in biology, medicine, information technology or social systems. In particular the application of the physical approach to social phenomena has been discussed in several reviews \cite{weidlich,SOOM,castellano}. As an example of the use of master equations in this field, we mention a recent paper \cite{Wio} in which the process of opinion formation in a society  is modeled as follows:
Society is divided in two parties, A and B, plus an ``intermediate`` group of undecided agents I. The supporters of A and B do not interact among them, but only through their interaction with the group I, convincing one of its members with a given probability. In addition there is a nonzero probability of a spontaneous change of opinion from I to the other two parties and vice-versa. More specifically, if $n_{A(B)}$ is the number of supporters of party A(B), $n_I$ is the number of undecided agents and $\Omega$ is the total number of individuals, the possible transitions are:

spontaneous change $A\rightarrow I$, occurring with a rate $\alpha_{1}n_A$,

spontaneous change $I\rightarrow A$, occurring with a rate $\alpha_{2}n_I$,

spontaneous change $B\rightarrow I$, occurring with a rate $\alpha_{3}n_B$,

spontaneous change $I\rightarrow B$, occurring with a rate $\alpha_{4}n_I$,

convincing rule $A+I\rightarrow 2A$, occurring with a rate $\frac{\beta_{1}}{\Omega}n_An_I$,

convincing rule $B+I\rightarrow 2B$, occurring with a rate $\frac{\beta_{2}}{\Omega}n_Bn_I$.

\noindent As the total number of individuals ($\Omega=n_A+n_B+n_I$) is fixed, there are only two independent variables, say $n_A$ and $n_B$. The master equation of the process is:
\begin{eqnarray}
&&\frac{\partial}{\partial t}P(n_A,n_B,t)=\alpha_{1}(n_A+1)P(n_A+1,n_B,t)+\alpha_{3}(n_B+1)P(n_A,n_B+1,t)\label{ME3}\\
&&+\alpha_{2}(\Omega-n_A-n_B+1)P(n_A-1,n_B,t)+\alpha_{4}(\Omega-n_A-n_B+1)P(n_A,n_B-1,t)\nonumber\\
&&+(\Omega-n_A-n_B+1)\left[\frac{\beta_{1}}{\Omega}(n_A-1)P(n_A-1,n_B,t)+\frac{\beta_{2}}{\Omega}(n_B-1)P(n_A,n_B-1,t)\right] \nonumber\\
&&-\left[\alpha_{1}n_A+\alpha_{3}n_B+(\alpha_{2}+\alpha_{4})(\Omega-n_A-n_B)+\frac{\beta_{1}n_A+\beta_{2}n_B}{\Omega}(\Omega-n_A-n_B)\right]P(n_A,n_B,t).\nonumber
\end{eqnarray}

We note that this master equation can be written in the general form (\ref{ME2v}) by setting $C_{1,0}=\alpha_{1}n_A$ , $C_{0,1}=\alpha_{3}n_B$ , $C_{-1,0}=(\Omega-n_A-n_B)(\alpha_{2}+\frac{\beta_{1}}{\Omega}n_A)$ and $C_{0,-1}=(\Omega-n_A-n_B)(\alpha_{4}+\frac{\beta_{2}}{\Omega}n_B)$.

An exact solution of this master equation is not known. In the following, we will apply to this problem the Gaussian approximation scheme and compare it with the results of the $\Omega$-expansion. The exact equations for the first moments are:

\begin{eqnarray}
\frac{d\langle n_A(t)\rangle}{dt}&=&-(\alpha_{1}+\alpha_{2}-\beta_{1})\langle n_A\rangle+\alpha_{2}(\Omega-\langle n_B\rangle)-\frac{\beta_{1}}{\Omega}\langle n_A^{2}\rangle-\frac{\beta_{1}}{\Omega}\langle n_An_B\rangle,\\
\frac{d\langle n_B(t)\rangle}{dt}&=&-(\alpha_{3}+\alpha_{4}-\beta_{2})\langle n_B\rangle+\alpha_{4}(\Omega-\langle n_A\rangle)-\frac{\beta_{2}}{\Omega}\langle n_B^{2}\rangle-\frac{\beta_{2}}{\Omega}\langle n_An_B\rangle,\\
\frac{d\langle n_A^{2}(t)\rangle}{dt}&=&(\alpha_{1}+\alpha_{2}(2\Omega-1)+\beta_{1})\langle n_A\rangle+\alpha_{2}(\Omega-\langle n_B\rangle)-2(\alpha_{1}+\alpha_{2}-\beta_{1}+\frac{\beta_{1}}{2\Omega})\langle n_A^{2}\rangle\nonumber\\
&&-(2\alpha_{2}+\frac{\beta_{1}}{\Omega})\langle n_An_B\rangle-\frac{2\beta_{1}}{\Omega}\langle n_A^{3}\rangle-\frac{2\beta_{1}}{\Omega}\langle n_A^{2}n_B\rangle,\\
\frac{d\langle n_B^{2}(t)\rangle}{dt}&=&(\alpha_{3}+\alpha_{4}(2\Omega-1)+\beta_{2})\langle n_B\rangle+\alpha_{4}(\Omega-\langle n_A\rangle)-2(\alpha_{3}+\alpha_{4}-\beta_{2}+\frac{\beta_{2}}{2\Omega})\langle n_B^{2}\rangle\nonumber\\
&&-(2\alpha_{4}+\frac{\beta_{2}}{\Omega})\langle n_An_B\rangle-\frac{2\beta_{2}}{\Omega}\langle n_B^{3}\rangle-\frac{2\beta_{2}}{\Omega}\langle n_An_B^{2}\rangle,\\
\frac{d\langle n_A(t)n_B(t)\rangle}{dt}&=&-(\alpha_{1}+\alpha_{2}+\alpha_{3}+\alpha_{4}-\beta_{1}-\beta_{2})\langle n_An_B\rangle+\alpha_{2}(\Omega\langle n_B\rangle-\langle n_B^{2}\rangle)\nonumber\\
&&+\alpha_{4}(\Omega\langle n_A\rangle-\langle n_A^{2}\rangle)-\frac{\beta_{1}+\beta_{2}}{\Omega}(\langle n_A^{2}n_B\rangle+\langle n_An_B^{2}\rangle).
\end{eqnarray}

Denoting by $A_1, A_2,B_1,B_2,C$ the Gaussian approximations to the moments $\langle n_A\rangle,  \langle n_A^2\rangle, \langle n_B\rangle, \langle n_B^2\rangle$ and the correlation $\langle n_An_B\rangle$, respectively, and using the results in table \ref{gaussmoments},  we obtain:
\begin{eqnarray}
\frac{dA_1}{dt}&=&-(\alpha_{1}+\alpha_{2}-\beta_{1})A_1+\alpha_{2}\Omega-\alpha_{2}B_1-\frac{\beta_{1}}{\Omega}A_2-\frac{\beta_{1}}{\Omega}C,\\
\frac{dB_1}{dt}&=&-(\alpha_{3}+\alpha_{4}-\beta_{2})B_1+\alpha_{4}\Omega-\alpha_{4}A_1-\frac{\beta_{2}}{\Omega}B_2-\frac{\beta_{2}}{\Omega}C,\\
\frac{dA_2}{dt}&=&(\alpha_{1}+\alpha_{2}(2\Omega-1)+\beta_{1})A_1+\alpha_{2}(\Omega-B_1)-2(\alpha_{1}+\alpha_{2}-\beta_{1}+\frac{\beta_{1}}{2\Omega})A_2\nonumber\\
&&-(2\alpha_{2}+\frac{\beta_{1}}{\Omega})C-\frac{2\beta_{1}}{\Omega}(3A_1A_2-2A_1^{3})-\frac{2\beta_{1}}{\Omega}(A_2B_1+2A_1C-2A_1^{2}B_1),\\
\frac{dB_2}{dt}&=&(\alpha_{3}+\alpha_{4}(2\Omega-1)+\beta_{1})B_1+\alpha_{4}(\Omega-A_1)-2(\alpha_{3}+\alpha_{4}-\beta_{2}+\frac{\beta_{2}}{2\Omega})B_2\nonumber\\
&&-(2\alpha_{4}+\frac{\beta_{2}}{\Omega})C-\frac{2\beta_{2}}{\Omega}(3B_1B_2-2B_1^{3})-\frac{2\beta_{2}}{\Omega}(B_2A_1+2B_1C-2B_1^{2}A_1),\\
\frac{dC}{dt}&=&-(\alpha_{1}+\alpha_{2}+\alpha_{3}+\alpha_{4}-\beta_{1}-\beta_{2})C+\alpha_{2}(\Omega B_1-B_2)\\
&&+\alpha_{4}(\Omega A_1-A_2)-\frac{\beta_{1}+\beta_{2}}{\Omega}\left[B_{1}A_{2}+B_{2}A_{1}+2(A_{1}+A_{2})C-2A_{1}^{2}A_{2}-2B_{1}^{2}B_{2}\right].\nonumber
\label{opgauss}
\end{eqnarray}

In van Kampen's expansion method, we define $\phi_{A(B)}, \xi_{A(B)}$ such that $n_{A(B)}=\Omega\phi_{A(B)}+\Omega^{1/2}\xi_{A(B)}$.

The equations for the macroscopic components are \cite{Wio}:
\begin{eqnarray}
 \frac{d\phi_{A}}{dt}&=&-\alpha_{1}\phi_{A}+[\alpha_{2}+\beta_{1}\phi_{A}](1-\phi_{A}-\phi_{B}),\\
 \frac{d\phi_{B}}{dt}&=&-\alpha_{3}\phi_{B}+[\alpha_{4}+\beta_{2}\phi_{B}](1-\phi_{A}-\phi_{B}),
\end{eqnarray}
and for the fluctuations:
\begin{eqnarray}
 \frac{d\langle \xi_{A}\rangle}{dt}&=&-[\alpha_{1}+\alpha_{2}+\beta_{1}(2\phi_{A}+\phi_{B})-\beta_{1}]\langle \xi_{A}\rangle-(\alpha_{2}+\beta_{1}\phi_{A})\langle \xi_{B}\rangle,\\
 \frac{d\langle \xi_{B}\rangle}{dt}&=&-[\alpha_{3}+\alpha_{4}+\beta_{2}(2\phi_{B}+\phi_{A})-\beta_{2}]\langle \xi_{B}\rangle-(\alpha_{4}+\beta_{2}\phi_{B})\langle \xi_{A}\rangle,\\
\frac{d\langle \xi_{A}^{2}\rangle}{dt}&=&-2\alpha_{1}\langle \xi_{A}^{2}\rangle-2(\alpha_{2}+\beta_{1}\phi_{A})(\langle \xi_{A}^{2}\rangle+\langle \xi_{A}\xi_{B}\rangle)+2\beta_{1}\langle \xi_{A}^{2}\rangle(1-\phi_{A}-\phi_{B})\nonumber\\
&&+\alpha_{1}\phi_{A}+(\alpha_{2}+\beta_{1}\phi_{A})(1-\phi_{A}-\phi_{B}),\\
\frac{d\langle \xi_{B}^{2}\rangle}{dt}&=&-2\alpha_{3}\langle \xi_{B}^{2}\rangle-2(\alpha_{4}+\beta_{2}\phi_{B})(\langle \xi_{B}^{2}\rangle+\langle \xi_{A}\xi_{B}\rangle)+2\beta_{2}\langle \xi_{B}^{2}\rangle(1-\phi_{A}-\phi_{B})\nonumber\\
&&+\alpha_{3}\phi_{B}+(\alpha_{4}+\beta_{2}\phi_{B})(1-\phi_{A}-\phi_{B}),\\
\frac{d\langle \xi_{A}\xi_{B}\rangle}{dt}&=&-(\alpha_{1}+\alpha_{3})\langle \xi_{A}\xi_{B}\rangle-(\alpha_{2}+\beta_{1}\phi_{A})(\langle \xi_{A}\xi_{B}\rangle+\langle \xi_{B}^{2}\rangle)-(\alpha_{4}+\beta_{2}\phi_{B})(\langle \xi_{A}\xi_{B}\rangle+\langle \xi_{A}^{2}\rangle)\nonumber\\ &&+(1-\phi_{A}-\phi_{B})(\beta_{1}+\beta_{2})\langle \xi_{A}\xi_{B}\rangle.
\end{eqnarray}

\noindent From those we can recover the original variables $n_{A(B)}(t)$.

In the next figures we compare the results coming from both approximations (obtained by numerical integration of the previous equations) and from simulations of the process using the Gillespie algorithm, for some representative values of the parameters and initial conditions. Again, the Gaussian approximation reproduces better the values for the average and the second moment whereas in this case both methods perform very similarly for the fluctuations and correlation.

\begin{figure}[h]
\centering
\includegraphics[scale=0.4,angle=0,clip]{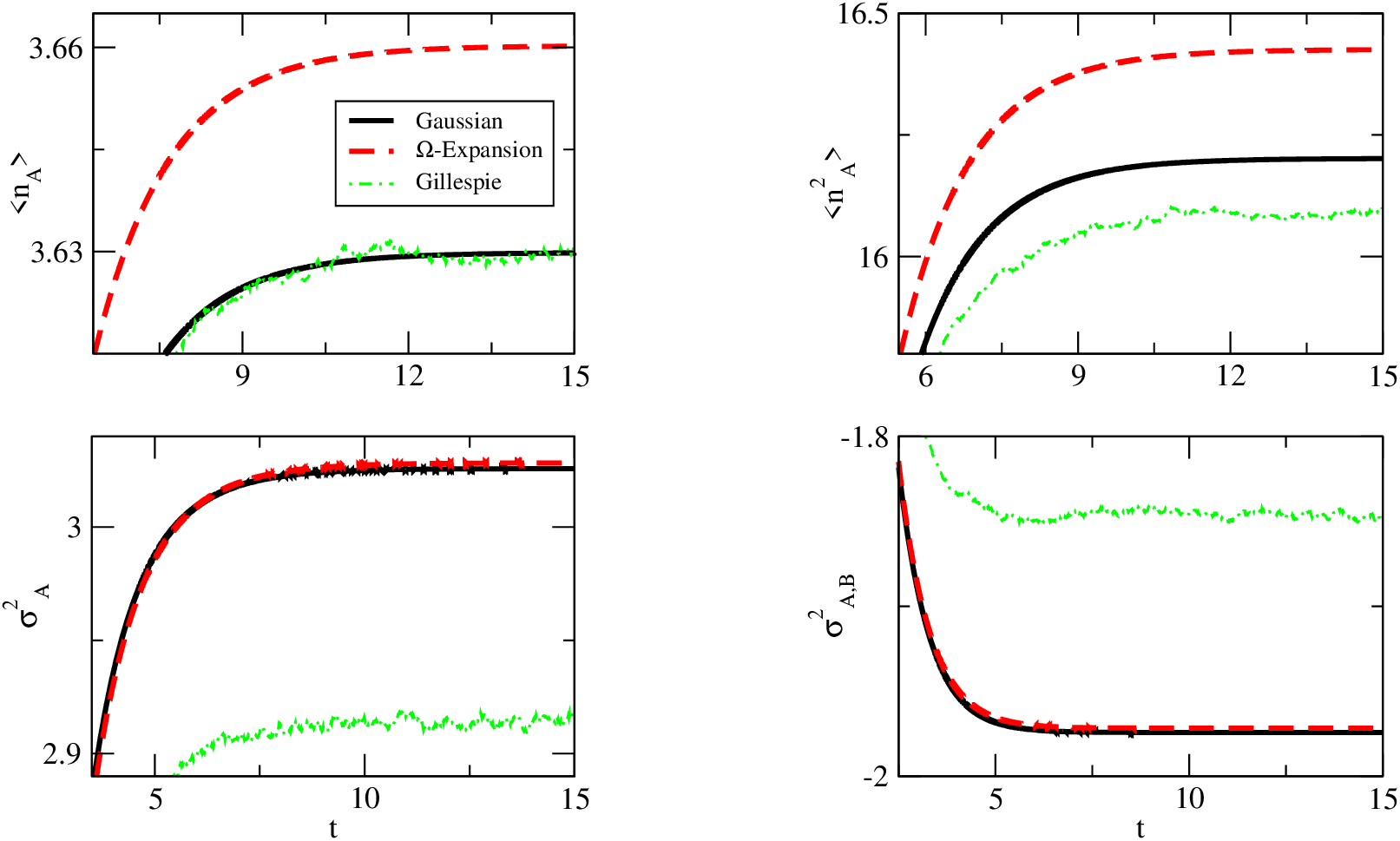}
\caption{$\langle n_A(t)\rangle, \langle n_A^{2}(t)\rangle$, $\sigma^{2}_{A}(t)$ and  $\sigma^2_{AB}(t)\equiv\langle n_{A}n_{B}\rangle-\langle n_{A}\rangle\langle n_{B}\rangle$ for the opinion formation model of reference \cite{Wio}, for  $\alpha_{i}=\beta_{i}=1, \Omega=10$, and initial conditions $n_A(0)=0,\,n_B(0)=\Omega$. For the average $\langle n_A(t)\rangle$, the Gaussian approximation (solid) follows very accurately the Gillespie simulation results (dot-dashed), whereas the $\Omega$-expansion (dashed) differs clearly. For the second moment $\langle n_A(t)^2\rangle$ the Gaussian approximation performs clearly better as well, while for the variance $\sigma^{2}_{A}(t)$ and  correlations $\sigma^2_{AB}(t)$, the Gaussian approximation and the $\Omega$-expansion give very similar results, although both are far from the simulation data.} 
\end{figure}

\section{Conclusions}
\label{sec:conclusions}
In this paper, we have given explicit expressions for the equations for the first and second moments of a stochastic process defined by a general class of master equations using the Gaussian approximation closure. The approach is motivated by van Kampen's $\Omega$-expansion result that, at lowest order, the fluctuations are Gaussian. We have shown that the Gaussian closure is simple to perform and leads to errors in the average value, the second moment and the fluctuations (the variance), that scale at most as $(\Omega^{-1/2},\, \Omega^{1/2},\,\Omega^{1/2})$, respectively. This is to be compared with the $\Omega$-expansion result in which the respective errors scale at most as $(\Omega^{0},\, \Omega^{1},\,\Omega^{1/2})$. Therefore, the Gaussian approximation is more accurate, which turns out to be important, specially for small values of $\Omega$. We have checked these results by comparing the performance of the two methods in three examples: (i) a binary chemical reaction, (ii) an autocatalytic reaction and (iii) a model for opinion formation. In all cases studied, the Gaussian closure has given a better approximation to the average and the second moment, although the $\Omega$-expansion, due to a cancellation of errors, yields a somehow smaller numerical error in the variance. In general, the Gaussian closure scheme is very simple to carry on in practice and this simplicity and the improvement of the predictive power is more apparent in many-variable systems. We believe that this method can be usefully applied to the study of other problems of recent interest in the literature involving stochastic processes in systems with a small number of particles.

\section{Appendix: Reaction-limited process}
\label{sec:solution}

We now find the solution of the master equation (\ref{eq:master}) in the equilibrium state for the general case, and the full dynamical solution for the irreversible case $\omega=0$.
Without loss of generality, let us rescale $t\to \kappa t/ \Omega$ and $\omega\to \omega \Omega^2/\kappa$ to get the simpler equation:
\begin{equation}
\frac{dP(n,t)}{dt}= (n+1)(\Delta +n+1)P(n+1,t)-n(n+\Delta)P(n,t) +\omega[P(n-1,t)-P(n,t)].
\end{equation}
Furthermore, only the case $\Delta\ge 0$ needs to be considered. If $\Delta<0$ the change $n'=n-\Delta$ leaves invariant the previous equation provided that we make the identification $P(n,t)\to P(n+\Delta,t)$. This means that the solutions in both cases are related by $P(n,t;\Delta)=P(n-\Delta,t;-\Delta)$.

The generating function 
\begin{equation}
\label{eq:fpn}
f(s,t)=\sum_{n=0}^\infty P(n,t)s^n,
\end{equation}
satisfies the partial differential equation:
\begin{equation}
\label{eq:pde}
\frac{\partial f}{\partial t}=(1-s)\left[s\frac{\partial^2f}{\partial s^2}+(1+\Delta)\frac{\partial f}{\partial s}-\omega f\right].
\end{equation}

Let us first discuss the equilibrium solution in the general case.

\subsection{The equilibrium solution}

By setting $\frac{\partial f}{\partial t}=0$ one gets the differential equation:
\begin{equation}
s\frac{\partial^2f}{\partial s^2}+(1+\Delta)\frac{\partial f}{\partial s}-\omega f=0.
\end{equation}
The solution around the singular regular point $s=0$ can be found by the Frobenius method as a power series $\sum_{n=0}^{\infty}a_ns^{n+\nu}$. The regular solution satisfying the boundary condition $f(s=1)=1$ is\footnote{There is another solution to this equation, but it contains a term in $\ln s$ and it has to be discarded since it can not be expanded in a power series of $s$. In the following $I_n(z)$ is the modified Bessel function of the first kind.}:
\begin{equation}
f(s)=\frac{s^{-\Delta/2}I_{\Delta}\left(2\sqrt{\omega s}\right)}{I_{\Delta}\left(2\sqrt{\omega s}\right)},
\end{equation}
and the equilibrium probabilities, rescaling back to the original parameters, are:
\begin{equation}
P(n)=\frac{(\omega\Omega^2/\kappa)^{n+\Delta/2}}{I_{\Delta}\left(2\Omega\sqrt{\omega/\kappa}\right)n!(n+\Delta)!},
\end{equation}
from where the first two moments can be computed as:
\begin{equation}
\label{eq:eqmom1}
\langle n\rangle = \frac{I_{\Delta+1}\left(2\Omega\sqrt{\omega/\kappa}\right)}{I_{\Delta}\left(2\Omega\sqrt{\omega/\kappa}\right)}\Omega\sqrt{\omega\kappa},\hspace{0.5cm}
\langle n^2 \rangle = \Omega^2\omega/\kappa -\frac{I_{\Delta+1}\left(2\Omega\sqrt{\omega/\kappa}\right)}{I_{\Delta}\left(2\Omega\sqrt{\omega/\kappa}\right)} \Delta\Omega  \sqrt{\omega/\kappa}.
\end{equation}

\subsection{The time-dependent solution}

We now study how the system relaxes towards equilibrium. We will restrict ourselves to the irreversible case $\omega=0$. This corresponds to the process $A+B\rightarrow 0$, inert. The partial differential equation (\ref{eq:pde}) can be solved by the technique of separation of variables by trying solutions of the form $f(s,t)=f_1(s)f_2(t)$. This leads to the pair of ordinary differential equations:
\begin{eqnarray}
s(1-s)f_1''+(1-s)(1+\Delta)f_1'+\lambda^2f_1 & = & 0,\\
f_2'+\lambda^2 f_2 & = & 0,
\end{eqnarray}
being $\lambda^2$ the constant arising from the method of separation of variables. The solution of the time dependent function is ${\rm e}^{-\lambda^2t}$ and the solution of the $s$-function is the hypergeometric function\footnote{There is another solution to the second-order differential equation. As before, this solution has to be discarded since it can not be expanded in powers of $s$.} $F(-\mu_1,\mu_2;\Delta+1;s)$. The explicit series is:
\begin{equation}
\label{eq:hyper}
F(-\mu_1,\mu_2;\Delta+1;s)=\sum_{n=0}^{\infty}\frac{(-\mu_1)_n(\mu_2)_n}{(\Delta+1)_n}\frac{s^n}{n!}.
\end{equation}
$(a)_n$ is the Pochhammer's symbol: $(a)_n=\frac{\Gamma(a+n)}{\Gamma(a)}$, or \makebox{$(a)_0=1$, $(a)_n=a(a+1)\dots(a+n-1)$} for $n>0$, and we have introduced
\begin{equation}
\mu_1=\frac{-\Delta+\sqrt{\Delta^2+4\lambda^2}}{2},\hspace{1.0cm}
\mu_2=\frac{\Delta+\sqrt{\Delta^2+4\lambda^2}}{2}.
\end{equation}
The solution for the function $f(s,t)$ is obtained by linear combination of the elementary solutions found above:
\begin{equation}
f(s,t)=\sum_{\lambda}C_{\lambda}F(-\mu_1,\mu_2;\Delta+1;s){\rm e}^{-\lambda^2 t}.
\end{equation}
This function is, in general, an infinite series on the variable $s$. In fact the coefficients, according to (\ref{eq:fpn}) are nothing but the time-dependent probabilities. However, in this irreversible case, the probability of having more $A$-molecules that the initial number at  $t=0$, say $M$, has to be zero. Therefore the series must be truncated after the power $s^M$. This implies that in the previous expression only hypergeometric functions that represent a polynomial in $s$ can be accepted. This is achieved by forcing $\mu_1=k=0,1,2\dots,M$, since the series (\ref{eq:hyper}) becomes then a polynomial of degree $k$. The condition $\mu_1=k$ is equivalent to the parameter $\lambda$ adopting one of the possible values $\lambda_k=\sqrt{k(k+\Delta)}$. Finally, noticing that $\mu_2-\mu_1=\Delta$, the solution can be written as:
\begin{equation}
f(s,t)=\sum_{k=0}^M \sum_{n=0}^kC_k(\Delta,M){\rm e}^{-k(k+\Delta)t} B_{n,k}(\Delta)s^n.
\end{equation}
The notation emphasizes that $C_k$ depends both on $\Delta$ and $M$ but $B_{n,k}$ depends only on $\Delta$:
\begin{equation}
B_{n,k}(\Delta)=\frac{(-k)_n(k+\Delta)_n}{n!(\Delta+1)_n }.
\end{equation}
All that remains is to impose the initial condition. We start with $M$ $A$-molecules at time $t=0$, such that $f(s,t=0)=s^M$. This implies that the coefficients $C_k$ must satisfy:
\begin{equation}
\sum_{k=n}^M B_{n,k}C_k=\delta_{n,M},
\end{equation}
for $n=0,1,\dots,M$. The solution starts by finding first $C_M=1/B_{M,M}$ and then proceeds backwards to find $C_{M-1},C_{M-2},\dots,C_0$ in a recursive manner. After some lengthy algebra, the result is:
\begin{equation}
C_k(\Delta,M)=(-1)^k\frac{2k+\Delta}{k+\Delta}\frac{(k+1)_{\Delta}}{\Delta!}\frac{(M-k+1)_k}{(M+\Delta+1)_k},
\end{equation}
(in the case $\Delta=k=0$ the correct interpretation of the undetermined expression is $C_0=1$). Going back to the original time variable, we now give the expression for the probabilities:
\begin{equation}
P(n,t)=\sum_{k=n}^M C_k(\Delta,M)B_{n,k}(\Delta){\rm e}^{-k(k+\Delta)\kappa t/\Omega}. 
\end{equation}
The normalization condition $\sum_{n=0}^MP_n(t)=1$ is verified with the help of the relation $\sum_{n=0}^kB_{n,k}=\delta_{k,0}$. The relation $\displaystyle \sum_{n=0}^knB_{n,k}=(-1)^kk\frac{\Delta!}{(k)_{\Delta}}$ (the indetermination arising when $\Delta=k=0$ must be resolved as $0$) helps to find the average of the number of particles:
\begin{equation}
\label{eq:mom1}
\langle n(t)\rangle= \sum_{k=1}^M(2k+\Delta)\frac{(M-k+1)_k}{(M+\Delta+1)_k}{\rm e}^{-k(k+\Delta)\kappa t/\Omega}.
\end{equation}
The second moment $\langle n(t)^2\rangle$ can be found with the help of Eq.(\ref{dndtd}) as \makebox{$\displaystyle \langle n(t)^2\rangle=-\frac{\Omega}{\kappa}\frac{d\langle n(t)\rangle}{dt}-\Delta \langle n(t)\rangle$}, or:
\begin{equation}
\langle n(t)^2\rangle= \sum_{k=1}^M(2k+\Delta)(k^2+(k-1)\Delta)\frac{(M-k+1)_k}{(M+\Delta+1)_k}{\rm e}^{-k(k+\Delta)\kappa t/\Omega}.
\end{equation}

\begin{acknowledgements}
Financial support from  Ministerio de Ciencia e Innovaci\'on (Spain) and FEDER (EU) grant FIS2007-60327
is acknowledged. L. F. L. is supported by the JAE-Predoc program of CSIC. We acknowledge fruitful discussions with H.S. Wio.
\end{acknowledgements}



\end{document}